\documentclass[11pt]{article}
\usepackage[T1]{fontenc}
\usepackage[utf8]{inputenc}
\usepackage{imfellEnglish}
\usepackage[british]{babel}
\usepackage{microtype}
\usepackage{geometry}
\usepackage{parskip}
\usepackage{nicefrac}
\usepackage{csquotes}
\usepackage[hidelinks]{hyperref}
\title{Historical Reflections on Interest Rates and the Emergence of the Yield Curve}

\author{Prof. Olivier Guéant}

\date{2026}

\begin{document}

\maketitle

\begin{abstract}

This text grew out of a historical introduction initially written for a study of interest rates in
cryptocurrency markets. The difficulty of defining a term structure for a currency without a
conventional bond market led naturally to a more fundamental question: under what historical
conditions does a yield curve become observable at all? Credit existed long before modern money, and
interest-bearing loans are documented as early as ancient Mesopotamia. For much of history, the
surviving evidence lacks the institutional features that facilitate reliable comparisons of interest
rates by maturity: standardised debt instruments, sufficiently homogeneous borrowers, regular
issuance over a range of maturities, observable market prices, and liquid secondary markets. We
trace the gradual emergence of these conditions from ancient Mesopotamia, Greece, and Rome, through
medieval and early modern Europe, to the development of modern sovereign debt markets in the
nineteenth and twentieth centuries.

\end{abstract}

\section*{Introduction: Interest Rates Are Ancient, Yield Curves Are Not}

Yield curves are so familiar in modern finance that it is easy to regard them as almost natural
objects. On trading floors, curves associated with government debt, interbank markets, swaps, or
corporate issuers are standard components of market data. They summarise borrowing conditions across
maturities and, with appropriate modelling assumptions, allow discount factors to be inferred for
valuing future cash flows. They also serve as basic inputs for the calibration of interest-rate
models.

Their construction relies on well-established mathematical techniques, although the resulting curve
depends on instrument selection and modelling assumptions. Given sufficiently many market prices of
bonds or other fixed-income instruments with different maturities, one may infer discount factors or
zero-coupon rates by bootstrapping and interpolate between observable maturities using splines or
other smoothing procedures; alternatively, the entire curve may be estimated through some form of
penalised nonlinear regression. These techniques are by now standard (see, for instance,
\cite{brigo2001interest}).

What is less often emphasised is that the possibility of performing such a calculation is itself the
result of a long historical process. A yield curve cannot be inferred merely because lending takes
place or because interest rates exist. It requires observable claims on sufficiently comparable
debtors at several maturities. In practice, it also benefits enormously from standardised
securities, repeated issuance, transparent market prices, and secondary markets in which those
securities can be traded before maturity. In that sense, a yield curve is very much an institutional
object.

The present note originated from a rather different question. In studying interest rates in
cryptocurrency markets, we were confronted with currencies for which active spot and derivatives
markets existed, but no conventional bond market offered a regular cross-section of fixed-rate debt
at different maturities. This made the usual construction of a yield curve impossible and raised a
natural historical question: is the simultaneous existence of a currency and of a yield curve in
fact the norm?

The answer is clearly no. Cryptocurrencies are unusual in many respects, but a monetary or economic
system in which credit exists without anything resembling a complete term structure of interest
rates is historically very ordinary. Indeed, interest-bearing credit is almost as old as written
economic records, whereas complete and continuously observable yield curves belong essentially to
the modern era.

The purpose of this note is therefore not to provide a history of credit in general, which would be
an immense undertaking, but to look at that history through a particular lens: \emph{when, and under
what conditions, does it become meaningful to speak about interest rates as a function of maturity?} This perspective leads to a simple distinction that will recur throughout the discussion. The
history of interest rates begins in Antiquity. The history of the yield curve begins much later.

The geographical scope of this note reflects the limits of my reading and
knowledge. My field is quantitative finance rather than history, and this
historical excursion already takes me beyond my usual area of expertise.
The narrative follows a predominantly Western trajectory, beginning with
ancient Mesopotamia and focusing thereafter on Europe and the United States.
It does not address the extensive trade and credit networks of the Islamic
world or the long monetary and financial history of China, subjects with
which I am insufficiently familiar to do them justice. These omissions
should not be taken to imply that the developments discussed here were
unique to the West or representative of financial history as a whole.

\section*{Ancient Credit Without a Yield Curve}

In the ancient Near East, Sumerian documents from the third millennium BCE attest to interest-bearing obligations involving grain measured by volume and silver measured by weight. The Cone of Entemena, dating to around 2400 BCE and housed at the Louvre, records the territorial dispute between Lagash and Umma, including obligations interpreted in terms of accumulated interest. Also housed at the Louvre, the famous Code of Hammurabi, circa 1750 BCE -- a Babylonian collection of legal provisions drawing on earlier Mesopotamian legal traditions -- explicitly specifies interest rate figures: 33\nicefrac{1}{3}\% for grain and 20\% for silver.\footnote{These
figures were probably chosen for their simplicity within the local system of fractional arithmetic
(see~\cite{hudson2002debt}).} These benchmark rates recur over many centuries in Mesopotamian
sources. Their legal status and application varied across periods and contracts, and their
interpretation requires attention to the period over which interest was charged. In practice, actual
interest rates could be lower, yet they still tended to be in double digits (see
\cite{sylla2013history}).

From the Babylonians, we can move to the Greeks. Coinage emerged in Lydia and the Greek cities of western Asia Minor toward the end of the seventh century BCE, before spreading more widely through the Greek world during the sixth century BCE. From maritime loans to loans secured by pledges, credit and debt played a
significant role in ancient Greek society, as evidenced indirectly by Solon's reforms in Athens. His
\textit{seisachtheia}, traditionally dated to 594/593 BCE, relieved existing debt burdens and prohibited loans secured on the debtor's person, releasing Athenians subjected to debt bondage.\footnote{Debt bondage
and debt remission appear in ancient texts; see Deuteronomy 15 for debt remission and Leviticus~25
for Jubilee, property restoration, and servitude.}

Regarding the characteristics of loans in classical Greece, loans to finance maritime trade were
very common (see~\cite{cohen1997athenian}) and could yield interest rates between 20\% and 30\% per
voyage (rather than per annum). Personal loans could be secured by movable assets (pledges) or by
land (\textit{hypotheke}) and, according to~\cite{sylla2013history}, typically had relatively short
durations, say up to a year or so, and involved monthly interest payments.

Loans to states or cities were relatively rare before the third or second century BCE and were
secured by guarantees from wealthy citizens, by hypothecated property, or by public revenues.\footnote{Sometimes, also, state loans were compulsory.} For example, in the fourth century
BCE, Demosthenes lent to the city of Oreus at 12\%, with the credit secured by the city's public
revenues (see~\cite{sylla2013history}).

Sources mention interest rates in the 16\%--18\% interval at the time of Solon, followed by a
gradual decline from 10\%--12\% in the fifth century BCE to 6\%--10\% by the end of the second
century BCE. Greek sources sometimes identify the purpose of a loan and its security, but the
surviving observations remain too heterogeneous to isolate the effect of maturity reliably.

Roman tradition associated the Twelve Tables, conventionally dated to 451--449 BCE, with
restrictions on interest. However, the numerical interpretation of the \textit{fenus unciarium} and
the subsequent history of Republican interest ceilings remain disputed. Many loans appear to have
been short-term, and real estate could serve as security (see~\cite{sylla2013history}).

The evidence becomes much richer from the first century BCE (see~\cite{andreau2001banque}). A
ceiling of 12\% is commonly reported for the late Republic, although the chronology and scope of the
relevant restrictions require caution. Prevailing interest rates could be well below this level,
especially during times of peace. As noted in~\cite{temin2004financial}, loans were sufficiently
widespread at the time to allow for general discussions about interest rates and standard lending
conditions. In his \textit{Letters to Atticus}, Cicero commented on an increase in interest rates
from \nicefrac{1}{3}\% to \nicefrac{2}{3}\% per month.

In the first century CE, normal interest rates in Rome typically ranged from 4\% to 6\%, except in
times of crisis, when they could rise to the legal maximum (and occasionally, though rarely, exceed
it). As in the previous century, loans were common: for instance, Columella's \textit{De re rustica}
(on agriculture) offers advice to those setting up vineyards, including how to manage costs related
to interest payments. Alongside lending by private individuals, professional bankers and money
changers performed a variety of financial services. Their activities included deposit-taking,
payments, and credit provision, although institutional arrangements varied across places and
periods.

The relatively low rates reported for the early Empire contrast with many surviving observations from subsequent centuries until the end of Antiquity.

All the available evidence suggests that nothing equivalent to a yield curve existed in antiquity. Lending practices appear to have offered too few comparable debts across maturities for even contemporary lenders and borrowers to construct one. This limitation seems to reflect the organisation of credit itself, not merely the gaps in our surviving sources.

\section*{Medieval Europe: Usury, Commerce, and the First Separation of Maturities}

The evolution of credit and interest rates in medieval Europe is inseparable from the Christian
doctrine on usury. In the Old Testament, we read in Deuteronomy 23:19--20:

\begin{quote}

``You shall not charge interest on loans to your brother, interest on money, interest on food,
interest on anything that is lent for interest. You may charge a foreigner interest, but you may not
charge your brother interest \dots'' (Deuteronomy 23:19--20, ESV)

\end{quote}

Christian writers increasingly interpreted the prohibition of usury as extending beyond lending
within a particular community. Ecclesiastical restrictions developed over centuries, initially
targeting clerics and subsequently imposing sanctions on lay usurers. The Council of Nicaea in 325
CE prohibited usury for clerics, and later ecclesiastical measures extended sanctions to laypeople.
As for state and civil law, the Capitularies of Charlemagne forbade lending at interest for both
clergy and laity. Enforcement was not always consistent, and the practical provision of credit
coexisted with religious and legal restrictions.

Commercial activity in early medieval Western Europe varied substantially across regions. Political
fragmentation and warfare disrupted some routes, while some maritime and long-distance exchange continued
through Mediterranean ports and northern trading centres, when possible. By the end of the tenth century and into
the eleventh, commerce gradually began to revive in Europe. Venice, which had never fully ceased
commercial activity, saw its role expand through agreements with Constantinople and exchanges with
the Muslim world. Other cities, especially in northern Italy and across northern Europe, also gained
commercial importance as they secured greater autonomy, resulting in the emergence of a merchant
class.

Commercial expansion accelerated during the eleventh and twelfth centuries. Wealthy merchants acted
as bankers by providing commercial loans with maturities often aligned with major trade fair dates
(such as the important Champagne fairs). These loans typically lasted a few months. Returns on
credit could be embedded in exchange transactions (early forms of bill of exchange). For
the late twelfth century,~\cite{sylla2013history} documents interest rates of around 20\% in Genoa
and between 10\% and 16\% in what is now the Netherlands.

The twelfth century also saw the emergence of wealthy families who could invest in new debt
instruments taking the form of life or perpetual annuities, secured by property. Such arrangements
could be treated as purchases of income rights rather than interest-bearing loans, although their
acceptability depended on their contractual form and remained subject to debate. Reported
annual payments amounted to around 8\%--10\% of the purchase price for certain annuity contracts
(see~\cite{sylla2013history}). Forced loans imposed by cities, such as the \textit{prestiti} in Venice,
constituted another form of long-term debt that would play an important role in the centuries to
come.

This period provides an early contrast between short-term commercial credit and long-term income
claims, rather than a term structure for comparable obligations. Short-term maturities reflected the
calendar of trade fairs, while perpetual debts provided an observable yield on an instrument with no
contractual maturity. However, the debtors at the two ends of the spectrum were not the same.

The thirteenth century was a period of major developments in scholastic thought, notably through the
work of St. Thomas Aquinas. Scholastic writers distinguished prohibited payment for the use of a
loan from compensation justified by particular circumstances. Aquinas, for example, accepted
compensation for an actual loss caused by lending, while rejecting a contractual charge for merely
prospective profits. These distinctions did not amount to a general acceptance of interest-bearing
lending.

This century also saw the flourishing of international trade, with important economic centres such
as Genoa, Bruges, Cologne, and, notably for the wool trade, London. The expansion of commerce was
supported by new financial innovations led by the Italians, especially the Florentines, who began to operate as bankers on a continental scale.~\cite{sylla2013history}
reports double-digit interest rates on certain deposits and on most loans during this period, with a
general downward trend over the course of the century. It also witnessed the expansion of annuities
and perpetual loans, which in some cases became tradable instruments. A notable example is the
Venetian \textit{prestiti} -- although associated with forced loans -- initially issued with a
nominal rate of 5\%.

The thirteenth century in some ways offers a first glimpse of what finance would come to resemble in the
early modern period. Of course, the fourteenth century brought severe demographic and economic
disruption, notably through the Black Death and prolonged warfare -- the Hundred Years' War would continue until the mid-fifteenth century. However, in spite of these events, the evidence on loans and interest rates during the fourteenth and early fifteenth centuries is more abundant than in previous periods. It notably includes a price chronicle of the Venetian \textit{prestiti}, providing one of the earliest extended bodies of price evidence
for the debt of a single public issuer.\footnote{When payments were suspended or uncertain, the
ratio of the contractual annual payment to the market price did not necessarily measure the return
investors expected to receive.} This evidence allows us to trace a long-term borrowing-cost
indicator for a single issuer, rather than the limiting yield of a zero-coupon term structure.

\section*{Early Modern Europe: From Public Debt to Organised Markets}

The period following the end of the Hundred Years' War was one of discoveries and inventions, during
which commerce once again expanded across Europe and would soon extend beyond. The fifteenth century
was also that of the Medici Bank and of several other banking institutions offering interest on
deposits (see~\cite{sylla2013history} for figures). During the second half of the century,
charitable lending institutions known as \textit{monti di pietà} were established in Italian cities
to offer an alternative to high-cost moneylenders. Interestingly, the low but nonzero rates charged
by these public institutions were defended as compensation for operating costs rather than usury,
although their legitimacy was initially debated.

During the first half of the sixteenth century, Antwerp, in the Habsburg Netherlands, emerged as a
leading European commercial and financial centre. Its Bourse served as a hub where bankers,
financial agents, and merchants exchanged credit instruments, including bills of exchange used in
international trade. Its position was then weakened by the political and military upheavals
of the Dutch Revolt.

Before the sixteenth century, princes were often heavily indebted and recurrently rolled over
short-term loans at rates higher than those for commercial credit, due to their poor
creditworthiness. In the case of the Habsburg government of the Low Countries,
\cite{sylla2013history} documents more than twenty short-term loans (most with maturities well below
one year) negotiated between 1509 and 1521, often in Antwerp or Bruges. Interest rates varied
considerably during this period: for example, an 18-month loan at 7\nicefrac{1}{2}\% in June 1510, a
4-month loan at 24\% in the same month, and a 10-month loan at 6\nicefrac{1}{4}\% less than a year
later. Clearly, despite the relatively high number of observations, there is no way to construct the
first segment of a yield curve in any consistent manner.

By the end of the sixteenth century, the model of long-term borrowing through life or perpetual
annuities had been widely adopted by cities such as Genoa,\footnote{The \textit{luoghi} in Genoa are
slightly different because payments were not fixed and should rather be regarded as dividends.}
Barcelona, and Amsterdam, as well as by states, including Francis I's France, which began issuing
\textit{rentes}, and the Spanish Crown, which refinanced its floating debt through perpetuals.
Fixed-maturity long-term loans remained very rare, but one notable example is that of the French
king Henri II, who refinanced the short-term rolling debt inherited from the wars of his father
(with Charles V, Holy Roman Emperor) by raising new funds through an 11-year loan in 1555, known as
the \textit{Grand Parti de Lyon}.

The sixteenth century also marked the beginning of a shift in perspective regarding usury, following
the Protestant Reformation. While Luther remained aligned with scholastic thinking on the subject,
this was not the case for Calvin, who argued that charging interest on commercial loans could be
acceptable. In England, legislation enacted under Henry VIII in 1545 permitted interest up to 10\%.
This measure was repealed in 1552, before a 10\% ceiling was restored under Elizabeth I in 1571 (see~\cite{munro2012usury}).

The seventeenth century was a century of contrasts.\footnote{We focus here on state borrowing, but
commercial loans were of course very common. One cannot speak of lending in early modern Europe
without thinking of Shylock, the moneylender of Shakespeare's \textit{The Merchant of Venice} --
whose famous bond, tellingly, carries no interest at all, only a pound of flesh as the penalty for
default. Merchants typically provided short-term credit directly to their suppliers and clients,
while long-term private lending often took forms such as mortgages and annuity contracts.} It was
both a century of expanding trade -- marked by the rise of major commercial companies and the
development of several exchanges across Europe, including the Amsterdam Bourse, which became a major
financial centre -- and a century of frequent wars that brought severe financial stress, including
several French defaults and repeated bankruptcies by the Spanish Crown, despite the wealth expected
to flow from the Americas.

In terms of credit structure, governments financed themselves both through short-term borrowing --
often at high interest rates -- and through very long-term instruments, as debt was increasingly
restructured in the form of perpetual loans (such as the \textit{rentes} in France), life annuities
(e.g., in the Dutch Republic and, at the very end of the century, in England), and some rare
long-term loans (exceeding 30 years in the case of Holland).

By the seventeenth century, interest rates were becoming increasingly national in character, with
particularly low rates (as low as 3\%) and minimal collateral requirements in Holland. Despite the
active presence of governments on both the short-term and very long-term segments, it remains
difficult to compute a clear yield curve for at least two reasons: (i) short-term rates were highly
volatile and appeared to be only weakly linked to default risk or the quality of collateral; and
(ii) although secondary markets existed for long-term debt, they had not yet developed for short-term instruments, making it difficult to track short-term rates consistently over time and relate them to those on longer-term obligations.

\section*{The Eighteenth Century: Standardisation and Geographic Differences}

London gained financial importance during the eighteenth century, while Amsterdam remained a major
international financial centre (see~\cite{ashton2013economic}). Both markets experienced speculative
episodes and financial crises. The Bank of England, founded at the very end of the seventeenth
century, played an important role in discounting bills. Its posted discount rate remained at 5\%
from 1746 to 1822; this stability should not be interpreted as evidence that market credit
conditions were equally stable.\footnote{From the middle of the eighteenth century, the financial
system increasingly financed industrial activity alongside commerce.} Decades later, it inspired the
creation of the \textit{Caisse d'Escompte} under Louis XVI in France, which issued notes and
extended short-term loans to the government at 4\%.

With regard to government borrowing, Britain borrowed recurrently during the eighteenth century,
mainly through perpetual loans and various types of annuities -- often incorporating lottery
features (see~\cite{cohen1953element}). Benefiting from declining interest rates, Britain refinanced
its debt by issuing consolidated annuities in the middle of the century, and soon introduced the 3\%
consols: perpetual loans with a 3\% coupon, redeemable at the issuer's discretion. Two major British
innovations -- standardisation and transparency in public finance -- contributed to the development
of an active secondary market. The prices of 3\% consols can be followed until 1888, when they were
exchanged for lower-coupon consols. For Britain, a continuous series of yields on a standardised
perpetual government obligation has thus been available since the mid-eighteenth century.

The French and Dutch cases differed in their institutional arrangements and the diversity of their
public debt instruments. French \textit{rentes} came in a wide variety of forms -- perpetuals, life
annuities, and even up to four-life annuities -- although many (issued at the beginning of the Seven
Years' War) carried a 5\% coupon.\footnote{Beyond financial market development, the disorder in
French public finance was so significant that the accumulation of debt over the eighteenth century
became one of the major contributing factors to the outbreak of the French Revolution in 1789.}
Dutch obligations also took various forms: there were multiple issuers beyond the Dutch Republic
itself, including towns and state-backed organisations, and a wide range of terms.
\cite{sylla2013history} documents perpetuals, life annuities, and fixed-term bonds of 30 and 32
years. The diversity of issuers and contractual forms complicated comparisons across securities, but
did not prevent secondary trading. Dutch public debt, in particular, was traded.

At the end of the eighteenth century, the United States gained independence from Great Britain and,
as early as 1776 and 1777, Congress authorised the issuance of medium-term domestic debt (3-year
maturity) at interest rates of 4\% and 6\%. The American Revolution and the early years of the
United States were also financed through loans obtained in Europe -- most notably from France and
the Netherlands -- with maturities ranging from 10 to 25 years. Notably, no perpetual debt was
issued during this early phase.

A foundational moment in the history of American public finance occurred with Hamilton's
restructuring of federal and state debts at the end of the century. As Secretary of the Treasury,
Hamilton implemented a funding programme that converted existing obligations into funded securities
with specified interest and redemption provisions. Together with measures to secure federal
revenues, it helped establish the federal government's creditworthiness.

\section*{The Nineteenth Century: The Curve Begins to Take Shape}

\subsection*{In Europe}

Throughout the nineteenth century, the vast majority of the British national debt consisted of
consols, and successive conversions of higher-coupon stock into 3\% consols reflected a sustained
effort toward standardisation. In 1888, taking advantage of persistently low interest rates and of
the government's right to redeem the debt at its discretion, Goschen converted the 3\% consols into
2\nicefrac{3}{4}\% consols, whose coupon was scheduled to fall to 2\nicefrac{1}{2}\% in 1903.

The nineteenth century also marked the beginning of the modern era in British public debt issuance,
with the growing use of instruments offering specified repayment horizons. The Exchequer bonds
proposed in 1853 carried declining coupons and provided for redemption after September 1894 under
specified holder or Treasury options. These instruments met with limited success. Far more popular
were the Exchequer bonds issued repeatedly in the late 1870s, this time with maturities up to 3 years. At
the short end, Treasury bills were introduced in 1877, typically with 3-month maturities and
occasionally with 6- or 12-month terms.

By the end of the century, a British yield curve -- albeit with only a few key points -- could thus
be reconstructed. It is particularly noteworthy that short-term interest rates were highly volatile
throughout the century, whereas long-term rates were far more stable and followed a structural
decline.

In France, \textit{rentes} remained extremely popular, although efforts toward standardisation were
less pronounced than in Britain. At various times, \textit{rentes} with 5\%, 4\nicefrac{1}{2}\%,
4\%, and 3\% coupons were all actively traded on the secondary market -- see \cite{sylla2013history}. Due to differing
probabilities of redemption, their yields often diverged, though they tended to follow the same
general trend as British consols -- albeit generally with a positive yield spread, potentially
reflecting differences in credit risk, liquidity, monetary conditions, and redemption
provisions.\footnote{During major political disruptions such as the revolutions of 1830 and 1848 and
the Franco-Prussian War of 1870, yield spreads increased significantly.} Interestingly, while
long-term rates were generally higher in France than in Britain in the series considered, the
reverse often held for short-term rates.

Dutch perpetuals and their yields exhibited similar patterns to their French counterparts: they
followed the trajectory of British consols, again generally with a positive yield spread, which
should not be attributed exclusively to credit risk, but with greater volatility. As in France,
Dutch short-term interest rates were generally lower than their British counterparts.

At the turn of the twentieth century, perpetual and other long-term funded obligations still
represented an important part of public debt in several European countries, alongside a growing
range of fixed-maturity instruments. This structure reflected a long institutional history involving
the legal treatment of annuities, established public borrowing practices, and investors' demand for
durable income streams.

What may seem most surprising is the enduring appetite of investors for such instruments given that perpetual bonds are, by design, highly sensitive to interest rate fluctuations. That said,
questioning investors' preference for these products might be anachronistic, given investors' demand
for stable nominal income and the characteristics of the alternatives available to them. Indeed, the
figure of the \textit{rentier} is a recurring one in the novels of Balzac and Zola, where characters
are described by the income generated by their annuities rather than by the market value of their
holdings.

\subsection*{Outside Europe}

In contrast to the pattern described above for several European countries, many loan contracts from
debtors outside Europe and in the British colonies -- often traded in London or Paris and
denominated in sterling or French francs -- included a maximum maturity date by which the principal
had to be repaid or incorporated amortisation features for gradual repayment. For a detailed list of
issues in London in the nineteenth century, see~\cite{sylla2013history}.

The case of the United States is special and deserves attention as the global financial centre would
shift from London to New York in the following century. During the nineteenth century, many issues
became redeemable at the government's option after a specified date. Although redemption often
occurred at or soon after that date, it remained contingent on refinancing conditions and should be
distinguished from contractual maturity. Over the first half of the century,\footnote{The federal
government retired its debt in 1835, but resumed borrowing after the financial crisis of 1837.} the
United States issued a series of bonds with periods until the earliest permitted redemption ranging
from 4 to 20 years.

A notable contractual feature appeared during the Civil War: bonds were issued with both an early
redemption date and a final due date, often with a substantial gap between the two. This structure
preserved the government's option to refinance while guaranteeing investors repayment by a specified
final date.

Following the Civil War and until the end of the nineteenth century, U.S. bond issuance remained
diverse, featuring early redemption dates corresponding to a wide range of periods until the
earliest permitted redemption. It is, however, difficult to
draw a genuine U.S. yield curve in the nineteenth century.

\section*{The Twentieth Century and the Birth of the Modern Yield Curve}

The twentieth century was marked by major geopolitical events -- including two world wars -- and
significant economic upheavals such as the Great Depression, which began in 1929, the high inflation
that followed the postwar boom, and the rise of a second wave of globalisation that continues into
the twenty-first century. It was also a century of profound technological transformation. The pace
of historical change accelerated to such a degree that any attempt to summarise the broader
historical context in just a few paragraphs would inevitably be reductive.

In terms of debt markets, it is possible to identify structural trends that broadened the maturity
coverage, liquidity, and regularity of sovereign debt markets. By the early 1940s, systematic yield-curve estimation was already well established, notably in studies of corporate bonds. Durand's
1942 study, for example, constructed yield curves using observations extending back to 1900 (see
\cite{durand1942basic}); this retrospective reconstruction should be distinguished from the
contemporaneous publication of curves. Postwar developments subsequently made sovereign yield curves
more widely and regularly observable.

Institutional investors assumed an increasingly important role during the twentieth century,
alongside individual investors. Their growing influence strengthened demand for instruments suited
to asset-liability management within banks, insurance companies, and pension funds. Together with
increased regulatory oversight and changes in public borrowing, this helped encourage the
diversification of bond maturities.

On the supply side, sovereign debt issuance surged due to the financing needs associated with the
two world wars and, subsequently, to persistent deficits driven in part by the expansion of welfare
states. Both France and the United Kingdom relied on a wide array of debt instruments -- not only in
terms of maturity but also with coupons indexed to inflation, gold, or foreign currencies -- to
maintain access to capital markets during challenging periods. In recent decades, regular issuance
across a broad range of maturities has become the norm, enabling the construction of sovereign yield
curves.

Perpetual securities gradually lost their central role in European sovereign borrowing as inflation,
changing investor preferences, and debt-management policies favoured other instruments. France's
remaining perpetual government \textit{rentes} were redeemed in the 1980s. British consols were finally
redeemed in 2015. Perpetual obligations thus became peripheral to modern sovereign debt markets.

In the United States, the creation of the Federal Reserve System helped stabilise short-term
interest rates, which had been extremely volatile in the nineteenth century. Regular issuance of
13-week Treasury bills began in 1929. Regular 26-week and one-year bills followed in 1959.
Nine-month bills, issued through reopenings of outstanding one-year bills, were introduced in
September 1966 and began to be phased out in September 1972.\footnote{See
\url{https://www.treasurydirect.gov/research-center/timeline/bills/} for the historical issuance
chronology.}

Beyond the short end of the curve, the United States issued a wide array of longer-dated instruments
throughout the century. Initially, many of these included both an early redemption date and a final
maturity date, often only a few years apart (especially from the 1930s onward). Over time, however,
the U.S. transitioned to issuing fixed-maturity bonds across the full maturity spectrum. From the
1970s onward, Treasury issuance became increasingly regular and structured, although the maturity
mix changed and issuance at particular maturities was sometimes suspended,\footnote{See
\url{https://www.treasurydirect.gov/research-center/timeline/notes/} and
\url{https://www.treasurydirect.gov/research-center/timeline/bonds/}.} helping U.S. yield curves
become a standard reference in global finance.

\section*{What Does It Take to Produce a Yield Curve?}

Seen over the very long run, the historical evidence suggests that the emergence of the yield curve
depended on several conceptually distinct conditions. Some are necessary for meaningful comparisons,
while others improve the reliability and frequency of observation.

First, there must obviously be credit. This condition was satisfied thousands of years ago.

Second, lending must take place at different maturities. This condition is also ancient.
Mesopotamian, Greek, and Roman contracts already had different durations, and medieval Europe
simultaneously supported short commercial credit and long-term annuity contracts.

But these two conditions are far from sufficient.

A third requirement is sufficient comparability across borrowers and contractual terms. Comparing a
three-month commercial loan to a merchant with a perpetual obligation of a city tells us little
about the pure effect of maturity. To speak meaningfully of a term structure, one would ideally like
to observe the same borrower -- or at least borrowers with very similar credit characteristics -- at
several horizons.

Fourth, the debt contracts themselves must possess some degree of standardisation. If every loan
contains different repayment clauses, embedded options, collateral, currencies, or guarantees, the
corresponding yields cannot be ordered by maturity without substantial adjustments.

Fifth, one needs contemporaneous prices or executable quotations. These may come from primary
issuance, secondary trading, or derivative markets. Active secondary markets greatly improve the
frequency and reliability of price discovery.

Finally, observations spanning a sufficiently broad range of maturities improve the identification
of the curve. Dense and regular issuance facilitates this process and reduces reliance on
interpolation assumptions. Once several benchmark instruments coexist, mathematics can interpolate
between them. But interpolation is the last step of the process!

This distinction is worth emphasising. Modern curve-construction techniques may create the
impression that a yield curve is produced by bootstrapping, spline interpolation, or econometric
fitting. Historically, the decisive work was done much earlier by governments, legislators,
financial intermediaries, exchanges, and investors. They created standardised debt instruments,
liquid markets, benchmark issuers, and a sufficiently broad spectrum of maturities. The mathematics
merely connects the points that institutions have made observable.

\section*{Statements}

\subsection*{Acknowledgements}

I would like to thank Quentin Archer, Philippe Bergault, Sébastien Bieber, Julien Pincet and Wenkai
Zhang for our discussions on the topic.

\subsection*{Declaration of Generative AI Use}

During the preparation of this manuscript, I alternated between various versions of Google's Gemini,
OpenAI's ChatGPT, and Anthropic's Claude to assist with drafting, editing, spell-checking, and
syntactic refinement, with a view to improving the readability and language quality of the paper: as
I am French, the models indeed offered me some prospect of raising my syntax towards that of His
Majesty's subjects. I also used these tools to help cross-check historical claims encountered in
books, including dates and accounts of events. Claims that remained insufficiently supported were
removed to make the text more robust. After using these tools, I reviewed and edited
the content as needed, perhaps reintroducing a few Gallicisms and misspellings along the way, and
I take full responsibility for the final content of the article.

\end{document}